\documentclass[preprint]{aastex}

\begin{document}

\title{The Origin and Distribution of Diffuse Hot Gas in Spiral Galaxy
 NGC 3184} 

\author{N. E. Doane \altaffilmark{1}, W. T. Sanders \altaffilmark{2},
E. M. Wilcots \altaffilmark{1}, M. Juda \altaffilmark{3} }

\altaffiltext{1}{Department of Astronomy, University of Wisconsin,
Madison WI, 53706. doane@astro.wisc.edu, ewilcots@astro.wisc.edu}

\altaffiltext{2}{Department of Physics, University of Wisconsin,
Madison WI, 53706.  sanders@wisp.physics.wisc.edu}

\altaffiltext{3}{Harvard-Smithsonian Center for Astrophysics, 60
Garden Street, Cambridge, MA 02138}

\pagebreak

\begin{abstract}

Deep {\it Chandra} exposures reveal the presence of diffuse X-ray
emission with a luminosity of $1.3\times10^{39}$ ergs s$^{-1}$ from the
spiral galaxy NGC 3184.  This appears to be truly diffuse thermal
emission distinct from the low-luminosity LMXB emission.  While the
unresolved emission from older LMXBs is more uniformly distributed
across the galaxy, the diffuse X-ray emission is concentrated in areas
of younger stellar populations and star forming regions.  The surface
brightness of the diffuse emission over the spiral arms is five times
greater than in off-arm regions, and eight times brighter in H~II
regions than in non-H~II regions.  Spectral fits to the diffuse
thermal emission are consistent with a low temperature component, T
$\sim1.5\times10^6$ K, plus a higher temperature component, T
$\sim5\times10^6$ K.

\end{abstract}


\section{Introduction}

Observations of the diffuse soft X-ray background
\citep{mccam83,mccam90} reveal the presence of several large bubbles
of hot few-million degree plasma within a few hundred parsecs, Sco-Cen
and Orion-Eridanus, in addition to the local bubble of one-million
degree plasma.  How pervasive these bubbles and the hot phase of the
interstellar medium (ISM) are within the Milky Way is not known,
primarily because we cannot observe the hot gas in our Galaxy beyond a
few hundred parsecs due to the large cross-section for soft X-ray
absorption by the neutral interstellar material.  Most models suggest
that there are hot bubbles scattered throughout the disk, having a
filling factor of order 10\%, with the Sun by chance inside one
\citep{slav89,slav92,slav93,ferr98}.

In the usual model for a diffuse hot gas bubble, a combination of
supernovae and stellar winds create shocks that force the ISM
gas into a dense shell of neutral material surrounding the SNe and
remaining hot stars \citep{smith01}.  The diffuse gas that is left
within the expanding bubble is heated by the shock to temperatures of
several million degrees. The ultimate fate of the hot gas within the
bubble is still not well understood.  The bubble may break out of the
disk, feeding the type of X-ray halos seen around NGC 891 and NGC 253
\citep{breg02, strick02}, or it may be pressure-confined to the disk
\citep{boulares90}.  In either case, the defining morphological
characteristic of a bubble of hot gas is a volume emptied of cool gas,
but containing hot gas emitting thermal X-rays, and typically
surrounded by a dense shell containing cooler shock-ionized and
neutral gas.  In this scenario diffuse X-ray emission in a galaxy will
be associated with recent SNe and hot stars, both signatures of
star-forming regions.  The ultimate fate of the bubble depends on its
energy (i. e. expansion velocity) and the conditions of its
surrounding environment (i. e., density and pressure distributions).

The question of the pervasiveness of the diffuse hot gas within a
galaxy cannot be well addressed by observations of our Galaxy, but can
be addressed in nearby face-on galaxies.  Absorption within our Galaxy
requires that we observe other systems to gain an understanding of the
distribution of the hot ISM on a galactic scale.  Until recently,
observations of diffuse X-ray emission in other spiral galaxies have
been limited by instrument angular resolution, as with the {\it
Einstein} and {\it ROSAT} observations.  The IPC instrument on {\it
Einstein} had an angular resolution of 1$^{\prime}$ and the PSPC
instrument on {\it ROSAT} had a 50\% encircled energy radius averaged
over the field of view of $1.'8$ \citep{asch88}.  Without better
angular resolution, point source emission can not be distinguished
from regions of diffuse thermal emission.  \citet{mccam84} used {\it
Einstein} to observe M101, and \citet{cui96} looked at five face-on
spiral galaxies, including M101, with {\it ROSAT}, but both were able
to put only upper limits on the total flux coming from the diffuse
emission.  Now with the {\it Chandra X-ray Observatory's} resolution
of $<$1$^{\prime\prime}$, point sources can be identified and removed,
making observations of the diffuse emission from other galaxies
feasible \citep{wang01,frat02,strick02,swartz03,kuntz03}.

We obtained deep X-ray and optical observations of NGC 3184 to
investigate the filling factor of the diffuse X-ray emission and to
determine the spatial correlation between star forming regions and
diffuse hot gas in a galaxy similar to the Milky Way.  Particulars of
NGC 3184 are given in Table 1.  In \S2 we present the data and details
of the X-ray and optical imaging observations and reductions.  Section
3 is devoted to the analysis of the point source population and the
spectral properties of the X-ray emission.  Our results are presented
in \S4.  Comparisons between the properties of the diffuse hot gas in
NGC 3184 and those of superbubbles in our Galaxy are discussed in \S5.

\section{Observations and Reductions}
\subsection{{\it Chandra} Data}

{\it Chandra} observations of NGC 3184 were made with the galaxy
centered on the back-illuminated S3 chip of the ACIS instrument: a
42.66 ks exposure on 8 January 2000 (obs ID 804) and a 24.04 ks
exposure on 3 February 2000 (obs ID 1520).  The second observation was
performed to observe SN1999gi \citep{sch01}, which occurred near the
center of NGC 3184 in December 1999.  We used the {\it Chandra} X-ray
Center's analysis program {\it CIAO} ({\it Chandra} Interactive
Analysis of Observations) version 2.3 and the {\it Chandra}
calibration files, CALDB Version 2.21, to reduce the observations.
Our reduction processing followed the guidelines released by {\it
Chandra} X-Ray Center (CXC) and utilized various of the {\it CIAO}
``science threads,'' which can be found at
http://asc.harvard.edu/ciao/documents\_threads.html.

Enhanced charged particle fluxes unrelated to the observed X-ray
source are often seen during observations with the S3 chip, and our
observations contained a number of them.  For a discussion of these
``flares" see \citet{pluc00}.  The flares were identified by
generating a light curve of events in the 2.5-7.0 keV range, using an
iterative-clipping algorithm to determine the quiescent mean rate, and
rejecting intervals that exceeded the mean-plus-three-sigma rate.
After filtering, the combined observations provided 56.50 ks of useful
data.  We re-gridded the two event files around a common tangent point
using the associated aspect solution files in the {\it CIAO} tool {\it
reproject\_events} so that the event lists could be combined using a
common coordinate system.  The data from the two observations were
then combined into a single image.  For our analysis we used only the
region where the two observations overlapped.

The full energy range that we used for analyzing our {\it Chandra}
observations is 0.36-8.0 keV.  Below 0.36 keV the increasing
electronic noise swamps the X-ray signal coming from the sky.  At
higher energies there is negligible instrumental response above 8.0
keV.  There is minimal diffuse emission above 1.3 keV from NGC 3184 so
we restrict our band pass to 0.36-1.3 keV for image analysis, but
maintain a larger band pass for point source detections and spectral
fits.  An image of the 0.36-1.3 keV diffuse soft X-ray emission is
shown in Figure 1.  Point sources have been removed via the procedures
described in \S3.1.  The raw image has a plate scale of 0.492 arcsec
pixel$^{-1}$, but the image in Figure 1 was smoothed using a six-pixel
($2.''95$) Gaussian filter.  There is an enhancement of the 0.36-1.3
keV band X-ray surface brightness over the face of NGC 3184.  The
decrease in signal off the galaxy can be better seen in Figure 2,
which shows a plot of counts s$^{-1}$ arcsec$^{-2}$ (in the band pass
0.36-1.3, with point sources removed) as a function of distance from
the galaxy's center.  The dashed line in Figure 2 indicates the level
of the non-X-ray instrument background, and the solid line on the
right hand side shows the level of the total background off-galaxy
--both were determined from the spectral fits described in \S3.2.  We
defined the X-ray extent of NGC 3184 by where the X-ray radial profile
flattens, at a radius of $3.'2$.  A circle of this radius was used to
separate the on-galaxy region from the off-galaxy region and is
indicated by the red circle in Figure 1.  A circle of radius $3.'2$ is
smaller than the optical definition of the galaxy given in Table 1.

\subsection{Optical Data}

We observed NGC 3184 over two nights in March and April of 2001 using
the WIYN{\footnote{The WIYN Observatory is a joint facility of the
University of Wisconsin-Madison, Indiana University, Yale University,
and the National Optical Astronomy Observatories.}}  3.5-m telescope
at Kitt Peak, AZ.  The WIYN observations consisted of two B band
exposures adding up to 2400 s, one 900 s V band exposure, two R band
exposures combining for 1200 s, three H$\alpha$ ($\lambda_{\circ}$ =
6565\AA, $\Delta\lambda~\simeq$ 47\AA) exposures for a total of 3600 s
and one H$\alpha$ continuum ($\lambda_{\circ}$ = 6618\AA,
$\Delta\lambda~\simeq$ 72\AA) exposure for 1200 s.  All observations
were made with the Mini-Mosaic camera, which has a $9.'6 {\times}
9.'6$ field of view and a 0.141 arcsec pixel$^{-1}$ plate scale.
Observing conditions were not photometric and therefore we were unable
to calibrate the optical images.

The Mini-Mosaic instrument is a combination of two CCD chips, each
with two amplifiers.  Each amplifier has a different bias, which
needs to be taken into account.  We corrected for the bias
difference, subtracted the over-scan (the background level from
the non-illuminated region of the chip), and flat-fielded the images
using dome flats.  To obtain adequate astrometry, we fit the
positions of 100 stars in the field and derived a plate solution.
We combined the data from the four amplifiers into a single image
for each exposure and removed the cosmic ray hits from each of those
images.  Finally we combined multiple observations that used the
same filter to create a single image in each band.  The B band image
of NGC 3184 is shown in Figure 3 and the continuum-subtracted
H$\alpha$ image is shown in Figure 4.  The horizontal strip in each
of these images is a result of a gap between the two CCD chips.

A radial surface brightness profile of the NGC 3184 B band image was
examined and shows that the extent of the galaxy is slightly smaller
in the X-ray than in the optical.  The B band image was used to define
the NGC 3184 spiral arms.  To define the spiral arms we used a series
of annuli, each centered on the center of the galaxy and $\sim$7$''$
wide, to generate azimuthal profiles from the B band image outside a
28$''$ radius circle.  Sections of the profiles that were 20\% greater
than the background-subtracted mean for that ring are used as a guide
to distinguish between the ``arm'' and ``off-arm'' regions of the
galaxy.  NGC 3184 has weak spiral arms, in that the brightness
contrast between ``arm'' and ``off-arm'' regions is low.  While this
is not a rigorous algorithm, it results in a good approximation of the
spiral structure of NGC 3184.  The central 28$''$ was also considered
part of the spiral arm.  Using this technique, we found 24\% of the
optical disk is ``arm.''  The ``arm'' region is outlined in blue in
Figures 1, 3 and 4.

The bright H~II regions were isolated using our H$\alpha$
observations.  We subtracted the 6618 \AA~image from the 6565
\AA~image to yield a continuum-subtracted H$\alpha$ image.  We
smoothed the image using a 31x31 pixel (4$.''$37) box filter and fit a
Gaussian to a histogram of the smoothed data.  Regions that were more
than 10 times brighter than the standard deviation of the Gaussian
distribution were identified as H~II regions.  Only 2.4\% of the
galaxy meets this criterion of emitting brightly in H$\alpha$.  These
regions are outlined in green in Figure 4.

Although neither the spiral arm region nor the  H~II regions are
defined using the X-ray data, the regions are correlated with the
distribution of X-ray emission.  The diffuse X-ray emission from NGC
3184 is not uniformly distributed across the galaxy.  Figure 2 shows
there is a radial dependence of the surface brightness of the hot gas
and we see X-ray emission out to $\sim3'$ ($\sim$10 kpc for a distance
of 11.6 Mpc), but the distribution is not simple.  In Figure 1, the
hot gas appears to have not only a radial dependence, but also to be
correlated with the galaxy's spiral arms.  Furthermore, where the
H$\alpha$ emission is bright, the X-ray emission is also bright, but
the converse is not true.  This is most noticeable in the central 1
kpc of the galaxy where there is bright X-ray emission, but negligible
H$\alpha$ emission.

\section{Analysis}

Before analyzing the diffuse X-ray emission, we have to account for
contributions to the flux from NGC 3184 from both point sources and
the background.  To do that we removed the resolved point sources and
modeled the remaining unresolved emission.  Section 3.1 addresses the
resolved point sources.  Section 3.2 addresses the non-X-ray and X-ray
backgrounds.  In section 3.3 we take into account the unresolved point
sources and fit the total unresolved emission from NGC 3184.

\subsection{Point Sources}

Point sources in the {\it Chandra} data were detected using the full
0.36-8.0 keV
 {\it Chandra} energy band and the {\it CIAO} program
{\it wavdetect} \citep{free02}.  {\it wavdetect} searches over
multiple scale sizes and correlates them to find point sources.  We
applied {\it wavdetect} using the logarithmic spatial scales of 1,
1.4, 2, and 4 pixels (chosen to optimize the wavelet search function),
and a significance threshold for source detection corresponding to a
less than 10$^{-6}$ chance probability of a false detection due to
local background fluctuations.  For each source, {\it wavdetect}
identifies an elliptical source region whose axes are three times the
standard deviation of the distribution of the source counts.  We
detected 61 sources in our 0.019 deg$^2$ field of view, 44 of them
from the 0.0096 deg$^2$ region on the face of NGC 3184 and 17 from the
0.0094 deg$^2$ region off of NGC 3184.  The latter number is roughly
consistent with that expected from the log N - log S function of
\citet{mus00}.  The locations of the 61 resolved sources are indicated
as green ellipses in Figures 1 and 3.

For each point source, the net number of detected counts in the
0.36-8.0 keV band within the source region defined by {\it wavdetect}
was determined by subtracting the local background normalized for each
source region.  The local background region was taken to be that
between concentric ellipses surrounding the source with axes 1.5 and 3
times the axes of the source ellipse.  To determine the flux from the
sources we used the spectral fit to the combined point source
spectrum, as described in \S3.3.  By dividing the number of counts in
the combined point source spectrum by the model-predicted flux in our
band pass, we determined a counts-to-flux factor that we used to
convert the number of counts from a given source to a flux.  To
correct for the differences in the instrument response across the
detector, we divided the flux from each point source by the relative
value of the exposure map at that location.  The exposure map was
created using the mean energy of the point source spectrum, 1.6 keV,
which was calculated using the spectral fits described in \S3.3.  The
exposure map combines for each pixel the exposure time, the effective
collecting area of the mirror, and the detector efficiencies and
geometry.  Assuming all 44 sources from the galaxy are at the distance
of NGC 3184, we converted source fluxes to luminosities, finding that
resolved sources have luminosities of $10^{37}$ to $10^{39}$ ergs
s$^{-1}$ in the 0.36-8.0 keV band.

The log N - log L (number vs. luminosity) relation for the 44 detected
sources from NGC 3184 is given in Figure 5.  This distribution is
similar to that found by \citet{kil02} in other nearby spiral galaxies
(including obs ID 804 of this NGC 3184 data set).  A one-component
power-law fit to the log N - log L data resulted in an unacceptable
reduced $\chi^2$ of 7.7.  We then used a broken power-law model to fit
the log N - log L data and found a reduced $\chi^2$ of 0.79.  The
break in the distribution occurs at $7.7\pm0.1{\times}10^{37}$ ergs
s$^{-1}$.  At higher luminosities the power-law exponent is
$1.09\pm0.01$, and for fainter luminosities the exponent is
$0.62\pm0.01$.  Following Kilgard (2002), the break in the source
luminosity function may be providing us with information on the last
episode of enhanced star formation in NGC 3184, but the change in
power-law index across the break, $\sim$0.5, is not very close to the
value of 1 expected in their model for an age-dependent luminosity
function.  The data start to fall away from the power-law fit at a
luminosity of $\sim2\times10^{37}$ ergs s$^{-1}$, indicating that below
this luminosity, our point source detection is probably incomplete.

\subsubsection{SN 1999gi}

One of the sources we removed from NGC 3184 was the Type II-plateau
supernova, SN 1999gi, reported by \citet{sch01}.  The position of SN
1999gi can be seen in the expanded B band image of Figure 6.  The
location of the SN is on the leading edge of the northern spiral arm,
in a region that contains a number of objects bright in the B band.
This is highly suggestive that its progenitor was an early-type star
in an OB association.  This region has an enhancement, independent of
the SN, of X-ray and H$\alpha$ flux (see Figure 1 and 4) and is
possibly a superbubble.

\subsection{Background Fitting}

To determine the surface brightness of NGC 3184, and the regions
within it, we performed spectral fits to the data from those regions
and used the best fit model parameters to characterize the surface
brightnesses.  Figure 7 shows the combined spectrum from the two
observations of NGC 3184, both on and off the galaxy, over the
0.36-8.0 keV energy range.  For spectral fitting, we used the energy
range of 0.36-3.0 keV so that we include significant higher energy
continuum.  In all of our spectral fits, we performed simultaneous
fits to the two independent data sets from the separate observations.
Although the sky spectra were required to be the same for the data
from the two independent observation dates, the normalizations of the
non-X-ray background, which is caused almost entirely by charged
particles, were allowed to be different for the two different dates.
All fits were performed using {\it Sherpa 3.0.2}.  We generated the
spectral response files, .rmf and .arf, for the 2000 February
observation, for which the detector temperature was -120$^{\circ}$ C,
using the {\it CIAO} program {\it acisspec}.  For the 2000 January
observation, when the temperature was -110$^{\circ}$ C, we used the
same .rmf used for the February observation because the calibration
files available from the {\it Chandra} website
(http://cxc.harvard.edu) for the -110$^{\circ}$ C time period are not
current.  We also applied the {\it apply\_acisabs} routine to the .arf
files to correct for the degradation in the ACIS quantum efficiency
over time (http://cxc.harvard.edu/cal/Acis/Cal\_prods/qeDeg/).
 
We used the off-galaxy portion of our data to determine the background
for the spectral analysis of the on-galaxy data.  The off-galaxy
background is made up of several components: local bubble emission,
Galactic halo emission, hot intergalactic medium (IGM), unresolved
AGNs, and the non-X-ray instrument background.  The hot IGM
contributes only about 10\% of the total background flux
\citep{phil01} so we did not include it as a separate component in our
model for the background.  The X-ray background has been studied in
previous X-ray missions, which provide the basis for the model
parameters we used to fit the X-ray background of our observations.
We adopted the temperatures found by \citet{sand02} from a
simultaneous fit to high latitude the X-ray background measured by
ROSAT \citep{snow97} and the Wisconsin all-sky survey \citep{mccam83}:
kT of 0.086 keV for the local bubble and 0.22 keV for the Galactic
halo.  All thermal models used in our fits are the Astrophysical
Plasma Emission Code (APEC, http://hea-www.harvard.edu/APEC) model.
To model the unresolved background AGNs we used a power-law with a
fixed index of 1.4 \citep{mccam90}.  Both the Galactic halo model and
the unresolved AGN model used ISM absorption \citep{mor83} equal to
that along the line of sight to NGC 3184 from Table 1.  To determine
the form of the non-X-ray background spectrum, we used the spectrum
from the {\it Chandra} Proposers' Observatory Guide
(http://cxc.harvard.edu/proposer/POG/index.html) shown in Figure 8.
This non-X-ray background, caused by charged particles interacting
with the {\it Chandra} satellite, does not pass through the telescope
in the same way X-ray emission does.  To properly treat the non-X-ray
background model, we used the {\it noise} command in {\it Sherpa} so
that it was not convolved through the telescope response function,
only the detector response is applied.  The non-X-ray background was
approximated using two Gaussians and a broken power-law.  The
Gaussians were used to fit the fluorescent emission lines,
Si-K$_{\alpha}$ at 1.74 keV and Au-M at 2.12 keV, caused by charged
particles interacting with the instrument.  For the broken power-law
we adopted a break energy of 0.7 keV, a low energy index of 1.47 and a
high energy index of 0.4.  In our spectral fits only the four
normalization parameters were allowed to float, those of the two
thermal models and of the two power-law models.  The spectral fit
parameters are given in the first section of Table 2.

After obtaining the best fit values for parameters of the background
model, we used those values to calculate the counts s$^{-1}$
arcsec$^{-2}$ of the X-ray background to compare with the image data
of Figure 2.  The value for the solid line on the right hand side of
Figure 2 was determined solely from the spectral fit parameters found
above.  As expected, the spectral fitting result agrees with the image
data.  As another consistency check, we compared the predictions of
our model parameters with the ROSAT data from around NGC 3184.  Using
the X-ray background tool from the HEASARC website
(http://heasarc.gsfc.nasa.gov/cgi-bin/Tools/xraybg/xraybg.pl), we
obtained the average ROSAT count rates in the R4, R5, R6 and R7 bands
over a 5$^{\circ}$ radius field of view centered on NGC 3184 that
excluded the inner 1$^{\circ}$ radius circle.  Our best fit model
predictions agreed with the ROSAT data to within 10\% in the 0.44-2
keV range.  Our observations of the X-ray background are also
consistent with the observed range of diffuse background surface
brightness measured with sounding rockets \citep{mccam90}, when
allowances are made for the different thresholds in the removal of the
resolved point sources from the AGN contribution.

 \subsection{Fitting Emission from NGC 3184}

The unresolved emission from NGC 3184 consists of two components,
unresolved point sources and truly diffuse emission.  The unresolved
point sources are composed of stars and the low luminosity end of the
low-mass X-ray binary (LMXB) population.  Assuming that the stellar
densities, X-ray luminosities and scale heights for X-ray emitting
stars given by \citet{kuntz01} for the Milky Way apply also to NGC
3184, we estimate that the unresolved emission from stars is 4\% of
the total unresolved emission.  The luminosities presented in
\citet{kuntz01} were calculated from ROSAT observations in the 0.1-2.4
keV range.  This energy range is larger than the 0.36-1.3 keV range we
use and results in 4\% being an upper limit on the contribution of
stellar emission.  A similar approach was taken by \citet{kuntz03} to
find the stellar contribution to the X-ray flux of M101 measured by
{\it Chandra} to be 10\%.  For the spectrum of the unresolved LMXBs,
we used the spectral form of the resolved point sources, but allowed
the normalization to float.  To determine the spectrum of the resolved
point sources we extracted the spectrum from each source and co-added
them to form a single spectrum.  This single combined spectrum of the
resolved point sources is best fit by an absorbed power-law with an
index 1.8 and absorption of $8.3\times10^{20}$ cm$^{-2}$.  This model
results in the same luminosity for the unresolved point source
emission as extrapolating the Log N - Log L plot of resolved point
sources down to $10^{35}$ ergs s$^{-1}$.  For the truly diffuse
emission we tried both one and two component models and found a
significantly better fit using the two thermal component model.

We then fit all of the on-galaxy data, varying only the normalization
of the point source power-law and the normalizations and temperatures
of the diffuse thermal components.  The off-galaxy background spectral
model was included as the on-galaxy background spectrum.  We found a
best fit ($\chi^2$ = 267 with 355 degrees of freedom) with kTs of
0.125$^{+0.034}_{-0.030}$ and 0.43$^{+0.25}_{-0.22}$ keV, and other
parameter values given in the second section of Table 2.  The best-fit
spectral model is shown in Figure 9 superimposed on the two data sets.
The error bars on the temperatures were taken from the one sigma
contour of the two temperature parameter space, shown in Figure 10 .
The two-temperature model for the diffuse X-ray emission in NGC 3184
is similar to that found by \citet{kuntz03} in M101, \citet{strick02}
in NGC 253, and \citet{frat02} in NGC 2403.  In subsequent fits to the
individual regions of NGC 3184, i.e. spiral arms, off-spiral arms,
bright H$\alpha$ regions, and non-H$\alpha$ regions, the temperatures
of the diffuse emission were frozen to the values obtained from the
fit to the whole galaxy due to lack of signal in the individual
regions.  The results of these fits are given in the latter sections
of Table 2.
 
\section{Results}

Using the results of the spectral fitting in the previous section, we
determined the surface brightnesses for these same regions of the
galaxy, as shown in Table 3.  We calculated the range of surface
brightnesses for each component using the model surface brightnesses
obtained from within the one sigma surface of the parameter space of
those parameters allowed to vary.  For the all-galaxy spectral fits,
the free parameter space consists of five dimensions.  The projection
of the volume within the one-sigma surface onto the
temperature-temperature plane is shown in Figure 10.  For the spiral
arm, off-arm, H$\alpha$, and non-H$\alpha$ spectral fits the parameter
space only involved three dimensions as the two temperatures were
fixed.  The resulting ranges in the surface brightnesses are shown in
Table 3.  We consider these surface brightnesses to be the primary
results of this investigation.

The surface brightness of the combined thermal emission of NGC 3184 in
the 0.36 - 1.3 keV band is $7.2\times10^{-19}$ ergs s$^{-1}$ cm$^{-2}$
arcsec$^{-2}$, which results in a luminosity of $1.3\times10^{39}$ ergs
s$^{-1}$ using the distance in Table 1 and the solid angle in Table 3.
The surface brightness of the point sources is $5.3\times10^{-19}$ ergs
s$^{-1}$ cm$^{-2}$ arcsec$^{-2}$.  The point source contribution is
relatively uniform across the galaxy, as can be seen in Table 3, where
each of the power-law surface brightnesses of the four sub-regions of
NGC 3184 is consistent with the all-NGC 3184 power-law surface
brightness.  This is the expected distribution throughout the galaxy
of the relatively old LMXB population.  The diffuse gas, however, is
clearly not uniformly distributed across the galaxy.  From Figure 1,
the total unresolved emission appears to be correlated with the
galaxy's spiral arms, and since the LMXBs are uniformly distributed,
the truly diffuse emission must not be.  From Table 3, the average
X-ray surface brightness of the combined thermal emission over the
optically-defined spiral arms is $1.9\times10^{-18}$ ergs s$^{-1}$
cm$^{-2}$ arcsec$^{-2}$, more than twice the average surface
brightness of the whole galaxy and a factor of 5 more than the X-ray
surface brightness of the ``off-arm'' regions.  Clearly the diffuse
hot gas tends to be spatially associated with the population of young
stars traced by what we call the spiral arms.

The diffuse X-ray emission is not only associated generally with the
spiral arms of NGC 3184, but also specifically with the H~II regions.
The surface brightness of the thermal component of the diffuse X-ray
emission coincident with H~II regions (see Figure 5) is
$4.8\times10^{-18}$ ergs s$^{-1}$ cm$^{-2}$ arcsec$^{-2}$, a factor of
7 higher than the X-ray emission not associated with H~II regions, as
shown in Table 3.  Our results are consistent with the picture that
the diffuse X-ray emission is broadly correlated with star-forming
regions.

\section{Discussion}

How does the emission we observe in NGC 3184 compare to that which we
observe in the Milky Way?  We can interpret the X-ray background in
our NGC 3184 data as a first approximation of the Milky Way X-ray
emission.  Seen externally, there are three components to the diffuse
X-ray signal coming from the solar neighborhood: (1) a Galactic halo
component equal to what we observed in the background fits toward NGC
3184, except with no absorption, (2) a local bubble component twice as
bright as what we saw, but with an absorption equal to that of our
Galaxy at high latitudes, and (3) a second Galactic halo component
(from the far side of our Galaxy) with absorption equal to twice what
we see at high latitudes.  This picture, using our emission measures
from Table 2, implies that we would measure a surface brightness from
the vicinity of the Sun of $8\times10^{-19}$ ergs s$^{-1}$ cm$^{-2}$
arcsec$^{-2}$ in the 0.36-1.3 keV energy range.  The surface
brightness of the local bubble component would be $5\times10^{-19}$
ergs s$^{-1}$ cm$^{-2}$ arcsec$^{-2}$ while the two halo components
add up to $3\times10^{-19}$ ergs s$^{-1}$ cm$^{-2}$ arcsec$^{-2}$.
This surface brightness estimate for the total emission is comparable
to the average surface brightness for all of NGC 3184 (see Table 3),
although it is only half of the average spiral arm surface brightness
measured in NGC 3184.  Thus, NGC 3184 appears to have diffuse X-ray
emission similar to what we see in the limited region of our Galaxy
that we can directly observe.

In NGC 3184 we see a correlation between regions of diffuse X-ray
emission and the spiral arms and H~II regions.  Over the course of
their short lifetimes massive stars deposit a tremendous amount of
mechanical energy into the surrounding interstellar medium, first in
the form of strong stellar winds and, subsequently, in the form of
supernovae explosions.  The energy injected into the ISM by massive
stars resides in both the thermal energy of the hot interior of the
bubble and in the kinetic energy of the expanding shell
\citep{McC87,Mac88}.  An individual massive star will deposit
$\sim2\times10^{51}$ erg of mechanical energy into the ISM in both a
stellar wind and a supernova explosion.  This is sufficient to heat
the interior of the bubble to X-ray emitting temperatures
\citep{Mac88}.  We therefore expect to find a strong spatial
correlation between the distribution of massive stars and the
distribution of bubbles of diffuse hot gas in galaxies.  For example,
\citet{dunne01} showed clearly that diffuse X-ray emission in the
Large Magellanic Cloud is associated with well-known regions of
massive star formation.

That the diffuse X-ray emission in NGC 3184 is associated
with discrete H~II regions suggests that the X-ray emission arises
from superbubbles associated with regions of star formation.  In our
Galaxy there are two nearby superbubbles centered on OB associations:
Orion-Eridanus and Scorpius-Centaurus (Sco-Cen).  The Sco-Cen
superbubble has a bright ridge, the North Polar Spur \citep{egger95},
so for this discussion we considered separately the North Polar Spur
and the interior of the Sco-Cen superbubble.  Observed parameters of
these regions are given in Table 4.  Using these observed parameters
in a thermal APEC model, we calculated the intrinsic surface
brightnesses of these regions, which are given in Table 5 along with
the surface brightnesses we determined from NGC 3184.  For
completeness we also include the surface brightness of the local
bubble as twice that seen along the line of sight to NGC 3184. 

Both the Orion-Eridanus superbubble and North Polar Spur have higher
surface brightnesses than the average NGC 3184 surface brightness, but
are comparable to that of the NGC 3184 spiral arms.  If the only X-ray
emitting gas in NGC 3184 were to come from Orion-Eridanus type objects
they could occupy no more than 30\% of the projected surface area of
NGC 3184 or less than 75\% of the spiral arms, otherwise we would have
detected more X-ray emission.  Similarly, the upper limit on how much
of the total NGC 3184 surface could be bubbles with the surface
brightness of the North Polar Spur is 50-100\%.  The surface
brightness of the interior of Sco-Cen is lower than, and that of the
local bubble is comparable to, the observed emission from NGC 3184.
If the entire galaxy were filled with either type of these bubbles we
would not be able to account for the diffuse X-ray emission seen from
NGC 3184.  Thus, the diffuse X-ray emission in NGC 3184 most likely
arises from a mixture of structures similar to those seen in the
Milky Way.

A potential concern when comparing NGC 3184 and the Milky Way is that
the brightest regions we observe in NGC 3184 are significantly
brighter than the brightest regions of diffuse thermal X-ray emission
in our Galaxy.  It is possible that this is a result of our location
in the Milky Way.  We can only study a small number of nearby
superbubbles because of the inherent difficulty of studying the X-ray
emission within our disk.  It is possible that the Milky Way contains
similarly bright regions to those of NGC 3184, but that we cannot
observe them.  Another concern is why we don't see in our observations
of NGC 3184 specific diffuse X-ray features of surface brightness
comparable to that of the nearby superbubbles in our Galaxy.  This is
most likely a result of {\it Chandra's} sensitivity and our short
observation time.  To produce one count in our 0.36-1.3 keV {\it
Chandra} observations requires a source luminosity of
$7.4\times10^{35}$ ergs s$^{-1}$.  To achieve a luminosity sufficient
to generate one count in our data, a superbubble with the temperature
and emission measure of the Orion-Eridanus bubble would need to
encompass 73 pixels (a 2$.^{\prime\prime}$4 radius circle).  Fainter
bubbles would need to be correspondingly larger: the interior of
Sco-Cen would have to comprise a 8$.^{\prime\prime}$7 radius circle.
One count, as generated by regions of these surface brightnesses,
would not be apparent in our observations because it would be spread
out across such a large detector area that it would be
indistinguishable from the noise.

The two temperatures that we observe from NGC 3184 straddle those
found in the superbubbles in our own galaxy.  The lower temperature
component has a temperature, $1.5^{+0.4}_{-0.3}\times10^6$ K, which is
warmer than our own local bubble, but cooler than the other
superbubbles that we see in our own Galaxy.  The high temperature
component, $5^{+2.9}_{-2.6}\times10^6$ K, is hotter than the nearby
superbubbles in our Galaxy.  This reinforces our conclusion that the
diffuse emission seen from NGC 3184 is a mixture of the types of
bubbles of our own Galaxy.  In the LMC, where individual superbubbles
have been detected and analyzed, temperatures in the range of
$1.7-9\times10^6$ K are found.  The diffuse X-ray emission from M101,
NGC 2403 and NGC 253 all have a similar lower temperature to the one
we find in NGC 3184, and high temperature components, T $\sim
8.0-9.0\times10^6$ K \citep{kuntz03,strick02,frat02}, consistent
within the upper limit of our measurement, although somewhat higher.
This is perhaps expected because these three galaxies all have
significantly higher 21 cm continuum flux than NGC 3184. Continuum
synchrotron radiation is an indicator of recent supernovae activity.
It is not unexpected that galaxies that have more recent star
formation have hotter gas that has not had time to cool.

In summary, we have detected diffuse X-ray emission arising from the
face-on spiral SABcd galaxy NGC 3184.  The power-law X-ray emission,
which is expected to come from unresolved LMXBs that have much longer
lifetimes than massive stars, is uniform across NGC 3184.  By contrast
the diffuse thermal emission, while observed throughout NGC 3184, is
strongly correlated with both the spiral arms and H$\alpha$-emitting
regions.  The diffuse thermal X-ray emission must be associated with
massive stars and star formation.  We conclude that because the
supperbubbles seen in NGC 3184 have surface brightnesses and
temperatures similar to what we observe in our Galaxy, the diffuse
thermal emission in NGC 3184 comes from a mixture of objects like we
see in our Galaxy.  The diffuse thermal emission across NGC 3184 is
pervasive and has an average surface brightness approximately equal to
that seen in our local neighborhood.  This suggests that the filling
factor of hot gas in NGC 3184, and possibly the Milky Way, may be
greater than the previously estimated 10\%.

\section{Acknowledgments}

We thank Andrew Glenn and Massimiliano Galeazzi for fruitful
discussions and the referee for useful comments.  We gratefully
acknowledge support from the National Science Foundation through grant
AST 00-98438, Chandra awards GO0-1022X and GO3-4113X, and funding from
NASA contract NASB-39073 to the {\it Chandra} X-ray Center.

\begin{table}
\scriptsize
\begin{center}
\begin{tabular}{|ccccc|}
\multicolumn{5}{c}{Table 1: NGC 3184 Particulars}\\
\hline
& \multicolumn{2}{c}{NGC 3184} & \multicolumn{2}{c}{Milky Way}\\
Parameter & Value & Source & Value & Source\\
\hline
Distance(Mpc) & 11.6 (56 pc arcsec$^{-1}$) & Leonard et al. (2002) & ... & ...\\
Position(J2000.0) & $\alpha$ = 10$^h$18$^m$17$^s$, $\delta$ = 41$^{\circ}$25$^{\prime}$28$^{\prime\prime}$ & ... & ... & ...\\
Galactic Coordinates & {\it l} = 178.33, {\it b} = +55.64$^{\circ}$ & ... & ... & ...\\
Foreground N$_H$(cm$^{-2}$) & 1.13$\times10^{20}$ & ... & ... & ...\\
Maj/Min diameter & 7$^{\prime}$.4/6$^{\prime}$.9 & ... & ... & ...\\
Hubble type & SAB(rs)cd & ... & Sb-Sbc &  Kerr (1993) \\
Star-formation rate(M$_\odot$ yr$^{-1}$ kpc$^{-2}$) & 10$^{-2.76}$ &
Larsen \& Richtler (2000) & 10$^{-2.40}$ & Tinsley (1980)\\
\hline
\end{tabular}
\end{center}
\end{table}

\begin{table}
\tiny
\begin{center}
\begin{tabular}{ccc}
\multicolumn{3}{c}{Table 2: Fit Parameters for NGC 3184 in the 0.36-3.0 keV range}\\
\hline
\hline
Parameters & Best Fit & $\chi^2$/dof \\
\hline
\multicolumn{3}{c}{Background}  \\
\hline
Non-X-ray broken power-law indices ...............................  & fixed at 1.5 \& 0.4 & 248/358\\
Non-X-ray broken power-law break energy$^a$ .................... & fixed at 0.7 &\\
Non-X-ray broken power-law normalization (obsID 804)$^b$  & 5700 & \\
Non-X-ray broken power-law normalization (obsID 1520) & 7300 &\\
AGN power-law index ...................................................... & fixed at 1.4 & \\
AGN power-law normalization$^c$ ........................................ & 5.8 & \\
AGN power-law absorption$^d$ ............................................. & fixed at 1.13 & \\
Thermal plasma (APEC) high temperature$^a$ ................... & fixed at 0.22 & \\
Thermal plasma high temperature emission measure$^e$ ..... & 0.0022 & \\
Thermal plasma high temperature absorption ................. & fixed at 1.13 & \\
Thermal plasma low temperature ..................................... & fixed at 0.086 & \\
Thermal plasma low temperature emission measure ........ & 0.027 & \\
Thermal plasma low temperature absorption ................... & fixed at 0 & \\
\hline
\multicolumn{3}{c}{All NGC 3184}  \\
\hline
Point-source power-law index ........................................... & fixed at 1.8 & 267/355 \\
Point-source power-law normalization .............................. & 11 & \\
Point-source power-law absorption ................................... & fixed at 8.3 & \\
Thermal plasma low temperature ..................................... & 0.125 & \\
Thermal plasma low temperature emission measure ......... & 0.016 & \\
Thermal plasma high temperature .................................... & 0.43 & \\
Thermal plasma high temperature emission measure ....... & 0.0016 & \\
Thermal plasma absorption .............................................. & fixed at 1.13 & \\
\hline
\multicolumn{3}{c}{Spiral Arms}  \\
\hline
Thermal plasma low temperature ..................................... & 0.125 & \\
Thermal plasma low temperature emission measure ......... & 0.016 & \\
Thermal plasma high temperature .................................... & 0.43 & \\
Thermal plasma high temperature emission measure ....... & 0.0016 & \\
Thermal plasma absorption .............................................. & fixed at 1.13 & \\
\hline
\multicolumn{3}{c}{Spiral Arms}  \\
\hline
Power-law normalization .................................................. & 5.9 &  170/357\\
Thermal plasma low temperature emission measure ......... & 0.035 &\\
Thermal plasma high temperature emission measure ....... & 0.0069 &\\
\hline
\multicolumn{3}{c}{Off-Spiral Arms}  \\
\hline
Power-law normalization .................................................. & 10 &  216/357\\
Thermal plasma low temperature emission measure ......... & 0.0097 &\\
Thermal plasma high temperature emission measure ....... & 0.00029 &\\
\hline
\multicolumn{3}{c}{Bright H$\alpha$}  \\
\hline
Power-law normalization .................................................. & 0 &  40.4/357\\
Thermal plasma low temperature emission measure ........ & 0.093 &\\
Thermal plasma high temperature emission measure ....... & 0.015 &\\
\hline
\multicolumn{3}{c}{Non-H$\alpha$}  \\
\hline
Power-law normalization ................................................. & 11 &  267/357\\
Thermal plasma low temperature emission measure ........ & 0.014 &\\
Thermal plasma high temperature emission measure ...... & 0.0013 &\\
\hline
\end{tabular}
\end{center}
a) keV

b) effective detector counts s$^{-1}$ keV$^{-1}$ sr$^{-1}$

c) photons s$^{-1}$ cm$^{-2}$ keV$^{-1}$ sr$^{-1}$ at 1 keV

d) 10$^{20}$ cm$^{-2}$

e) cm$^{-6}$ pc
\end{table}

\begin{table}
\scriptsize
\begin{center}
\begin{tabular}{|lcccccc|}
\multicolumn{7}{c}{Table 3: Surface Brightnesses$^a$ of NGC 3184 field
  in 0.36-1.3 keV}\\
\hline
& Solid Angle$^b$ & \multicolumn{3}{c}{Diffuse Thermal Emission} & Power-Law$^c$ & Total\\
Region & & 0.125 keV & 0.43 keV & Total &  &\\
\hline
X-ray Background & 110000 & & & \phn4.7$^{+20\%}_{-20\%}$ & 2.4$^{+\phn52\%}_{-\phn32\%}$ & \phn7.1$^{+19\%}_{-12\%}$ \\
All NGC 3184 & 114000 & \phn5.6$^{+20\%}_{-20\%}$ & \phn1.6$^{+\phn54\%}_{-\phn55\%}$ & \phn7.2$^{+20\%}_{-20\%}$ & 5.3$^{+\phn26\%}_{-\phn30\%}$ & 13.\phn$^{+11\%}_{-11\%}$\\
Spiral Arm & \phn27800 & 12.\phn$^{+17\%}_{-17\%}$ & \phn6.8$^{+\phn29\%}_{-\phn30\%}$ & 19.\phn$^{+14\%}_{-14\%}$ & 2.8$^{+110\%}_{-100\%}$ & 22.\phn$^{+11\%}_{-12\%}$\\
Off-Arm & \phn86700 & \phn3.4$^{+26\%}_{-25\%}$ & \phn0.3$^{+260\%}_{-100\%}$ & \phn3.7$^{+32\%}_{-32\%}$ & 4.9$^{+\phn28\%}_{-\phn29\%}$ & \phn8.6$^{+11\%}_{-13\%}$\\
H$\alpha$ & \phn\phn2190 & 33.\phn$^{+29\%}_{-38\%}$ & 15.\phn$^{+\phn52\%}_{-\phn91\%}$ & 48.\phn$^{+27\%}_{-39\%}$ & 0.0$^{+21.4}_{-\phn0.0\phn}$ & 48.\phn$^{+34\%}_{-30\%}$\\
Non-H$\alpha$ & 112000 & \phn5.0$^{+14\%}_{-16\%}$ & \phn1.3$^{+\phn56\%}_{-\phn53\%}$ & \phn6.3$^{+15\%}_{-16\%}$ & 5.2$^{+\phn22\%}_{-\phn24\%}$ & 12\phn$^{+\phn8\%}_{-\phn9\%}$\\
\hline 
\end{tabular}
\end{center}
a) 10$^{-19}$ ergs s$^{-1}$ cm$^{-2}$ arcsec$^{-2}$

b) arcsec$^2$, excluding the ellipses surrounding the resolved point sources

c) represents the AGN contribution for the background and the LMXB contribution for the galactic emission

\end{table}

\begin{table}
\begin{center}
\begin{tabular}{|ccccc|}
\multicolumn{5}{c}{Table 4: Nearby Regions}\\
\hline
& n$_e$ & Extent$^a$ & Emission Measure & T \\
Superbubble & (cm$^{-3}$) & (pc) & (cm$^{-6}$pc) & (10$^6$ K)\\
\hline
Orion-Eridanus & 0.015$^b$ & 140$^b$ & 0.032 & 3.3$^b$ \\
North Polar Spur & 0.01-0.02$^c$ & \phn60$^c$ & 0.01-0.02$^c$ & 3$^c$ \\
Sco-Cen interior & 0.0025 & 320$^d$ & 2.0 x 10$^{-3}$ & 4.6$^d$\\
\hline
\end{tabular}
\end{center}
a) distance through the emitting volume

b) Guo et al. (1995)

c) Iwan (1980)

d) Egger \& Aschenbach (1995)

\end{table}

\begin{table}
\begin{center}
\begin{tabular}{|cr|}
\multicolumn{2}{c}{Table 5: Surface Brightnesses in 0.36-1.3 keV band}\\
\hline
& Surface \\ 
Region & Brightness$^a$ \\ 
\hline
Orion-Eridanus & 26.\phn\\
North Polar Spur & 8.0-16.\\ 
Sco-Cen interior & 1.9\\  
Local Bubble & 5.3\\
Milky Way Halo$^b$ & 3.12\\
All NGC 3184 & 7.2\\
Spiral Arms in NGC 3184 & 19.\phn\\ 
Off-Spiral Arms in NGC 3184 & 3.7\\  
Bright H$\alpha$ in NGC 3184 & 48.\phn\\ 
Non-H$\alpha$ in NGC 3184 & 6.3\\ 
\hline
\end{tabular}
\end{center}
a) 10$^{-19}$ ergs s$^{-1}$ cm$^{-2}$ arcsec$^{-1}$

b) seen from outside the Milky Way, as described in \S 5
\end{table}

\centerline{\bf{Figure Captions}}

\figcaption{} 
{\it Chandra} observations of NGC 3184 in the 0.36-1.3
keV range.  The data were smoothed with a six-pixel Gaussian filter
and are displayed on a linear grey-scale with white values indicating
less than 0.04 counts arcsec$^{-2}$ and the black regions greater than
0.12 counts arcsec$^{-2}$.  The green ellipses indicate the locations
and the ellipse size represents the count distribution of the removed
point sources.  The red circle indicates the $3.'2$ radial extent
of the galaxy as determined from the X-ray data, while the spiral
arms used in the reduction, determined from the B band data,
are outlined in blue.

\begin{figure}
\epsscale{1.0}
\plotone{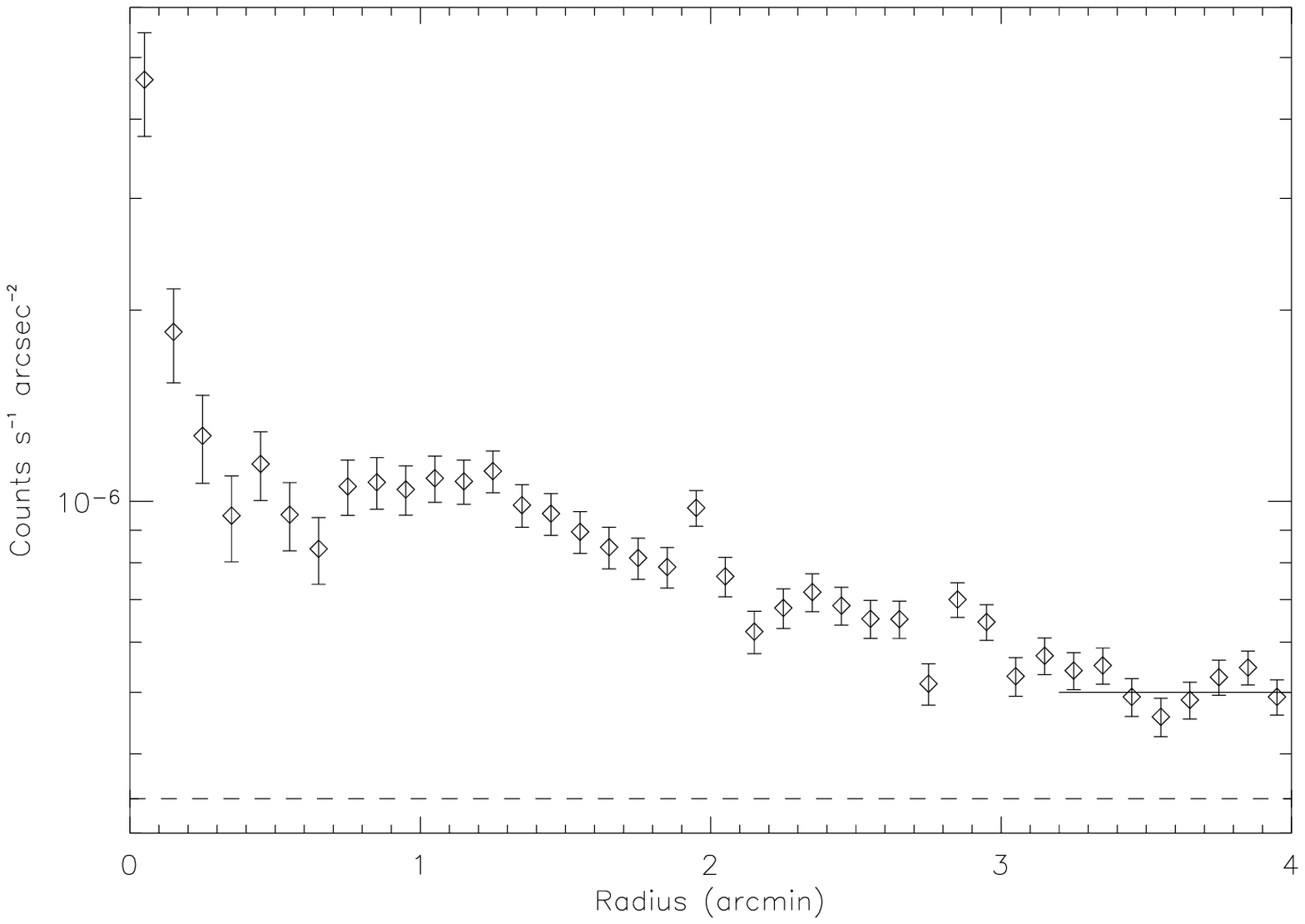}
\caption{
Radial brightness plot for raw counts from the NGC 3184 observations
in the 0.36-1.3 keV energy band with point sources removed.  The
dashed line denotes the level of non-X-ray instrument signal across
the detector.  The solid line represents the total background signal
coming from outside the $3.'2$ radial extent of NGC 3184, as
determined from spectral fitting described in \S3.2.}
\end{figure}

\figcaption{} 
The WIYN B-band image of NGC 3184.  The green ellipses indicate the
location and distribution of the X-ray point sources found using {\it
wavdetect} in the 0.3-8.0 keV band.  The red circle indicates the
$3.'2$ radial extent of the galaxy as determined from the X-ray data,
which is consistent with the surface brightness fall-off found in the
B band.  The area outlined in blue is the spiral arms region used in the
reduction, which were determined from the B band data.  The
horizontal strip is a result of a gap between the two CCD chips in the
Mini-Mosaic camera.

\figcaption{}
The WIYN H$\alpha$ continuum-subtracted image of NGC 3184.  The green
polygons mark regions of high H$\alpha$ emission.  The red circle
indicates the $3.'2$ radial extent of the galaxy as determined from
the X-ray data, while the spiral arms used in the reduction,
determined from the B band data, are outlined in blue.  The horizontal
strip is a result of a gap between the two CCD chips in the Mini-Mosaic
camera.

\begin{figure}
\epsscale{1.0}
\plotone{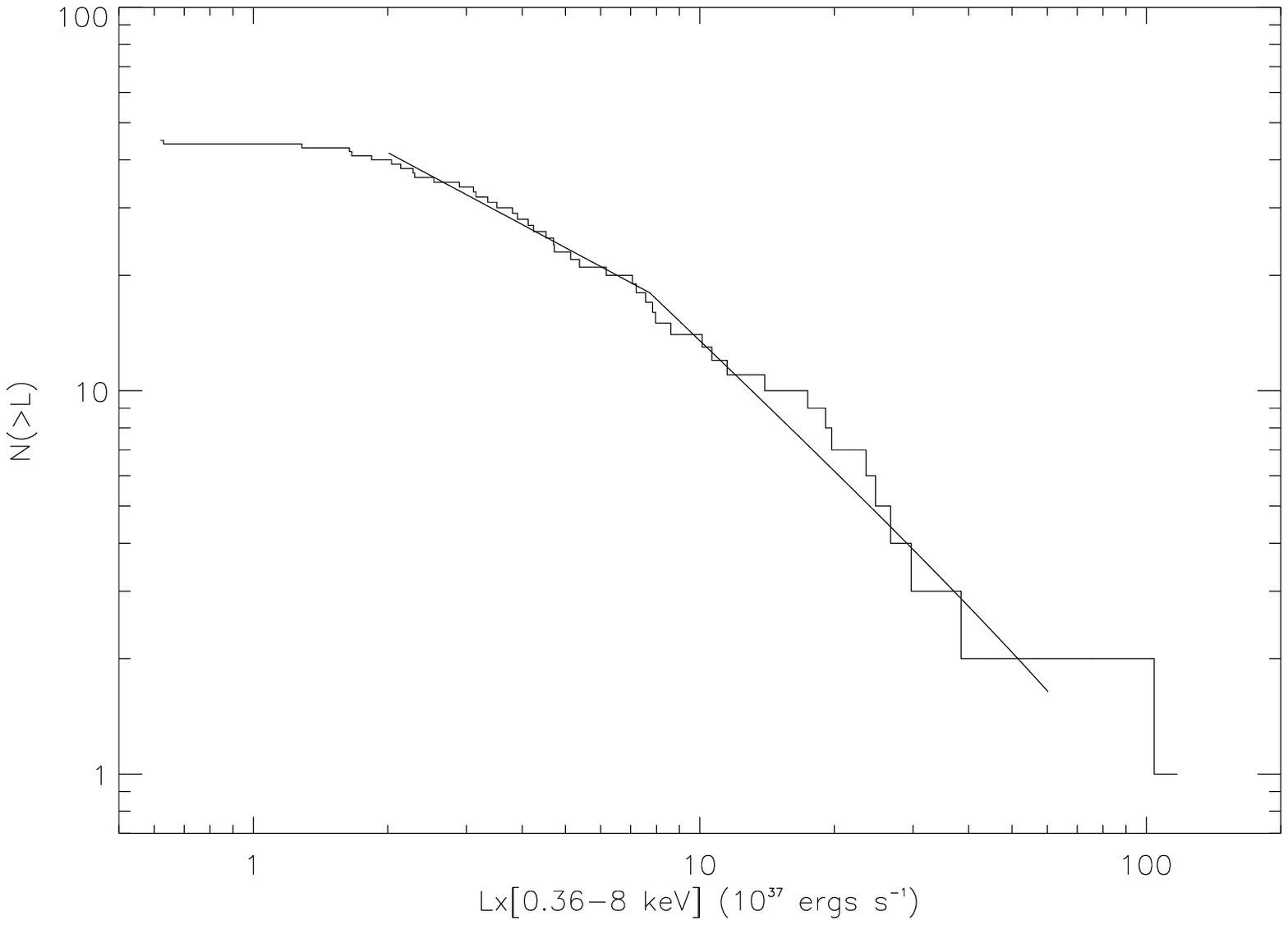}
\caption{Cumulative source luminosity function of NGC 3184 in the
0.36-8.0 keV energy band, for sources located within the 3.$'$2 radius
of the galaxy.}
\end{figure}

\figcaption{}
Expanded B band image of NGC 3184 with the position of SN 1999gi
indicated by the cross-hairs.

\begin{figure}
\epsscale{1.0}
\plotone{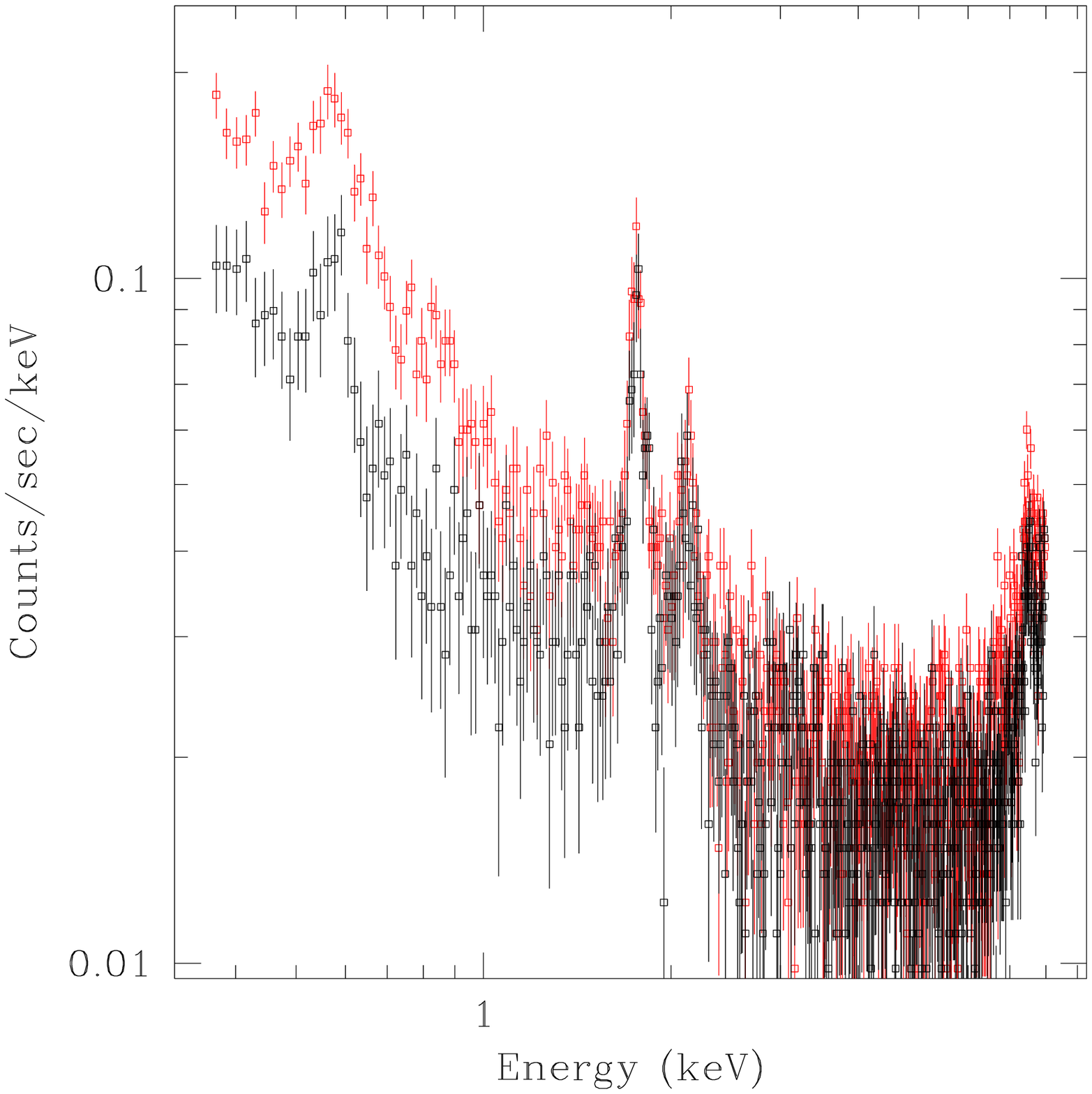}
\caption{The combined spectrum from the two NGC 3184 observations
in the full 0.36-8.0 keV energy band.  The red squares indicate data
from within the $3.'2$ radial extent of the galaxy.  The black squares
denote the off-galaxy spectrum.}
\end{figure}

\begin{figure}
\epsscale{1.0}
\plotone{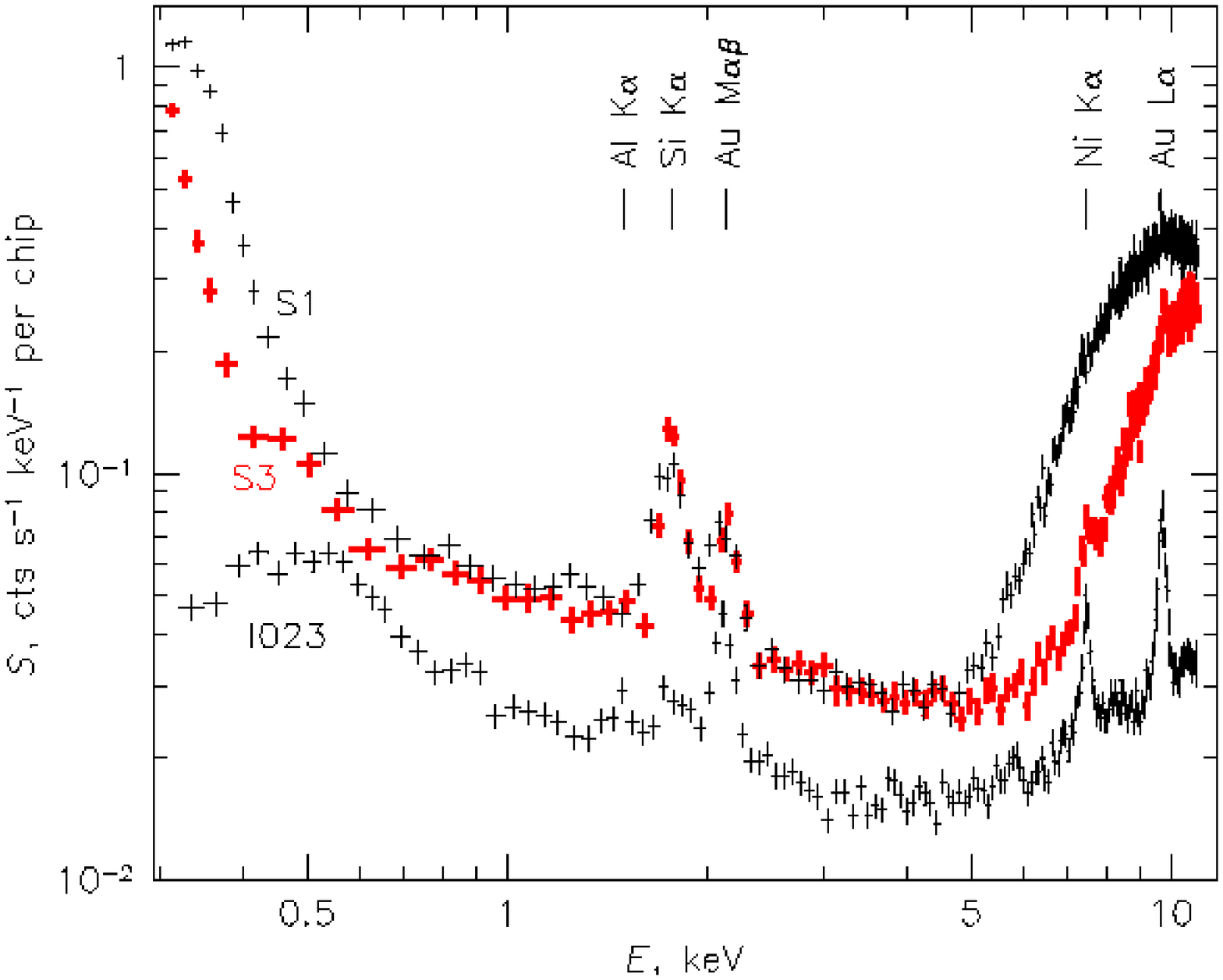}
\caption{2002 September ACIS observation obtained while in the stowed
position used to determine the non-X-ray background coming from
charged particles.  The emission lines are due to fluorescence of the
material in the telescope and focal plane.  This is figure 6.21 from
the {\it Chandra} ``Proposers' Observatory Guide."}
\end{figure}

\begin{figure}
\epsscale{1.0}
\plotone{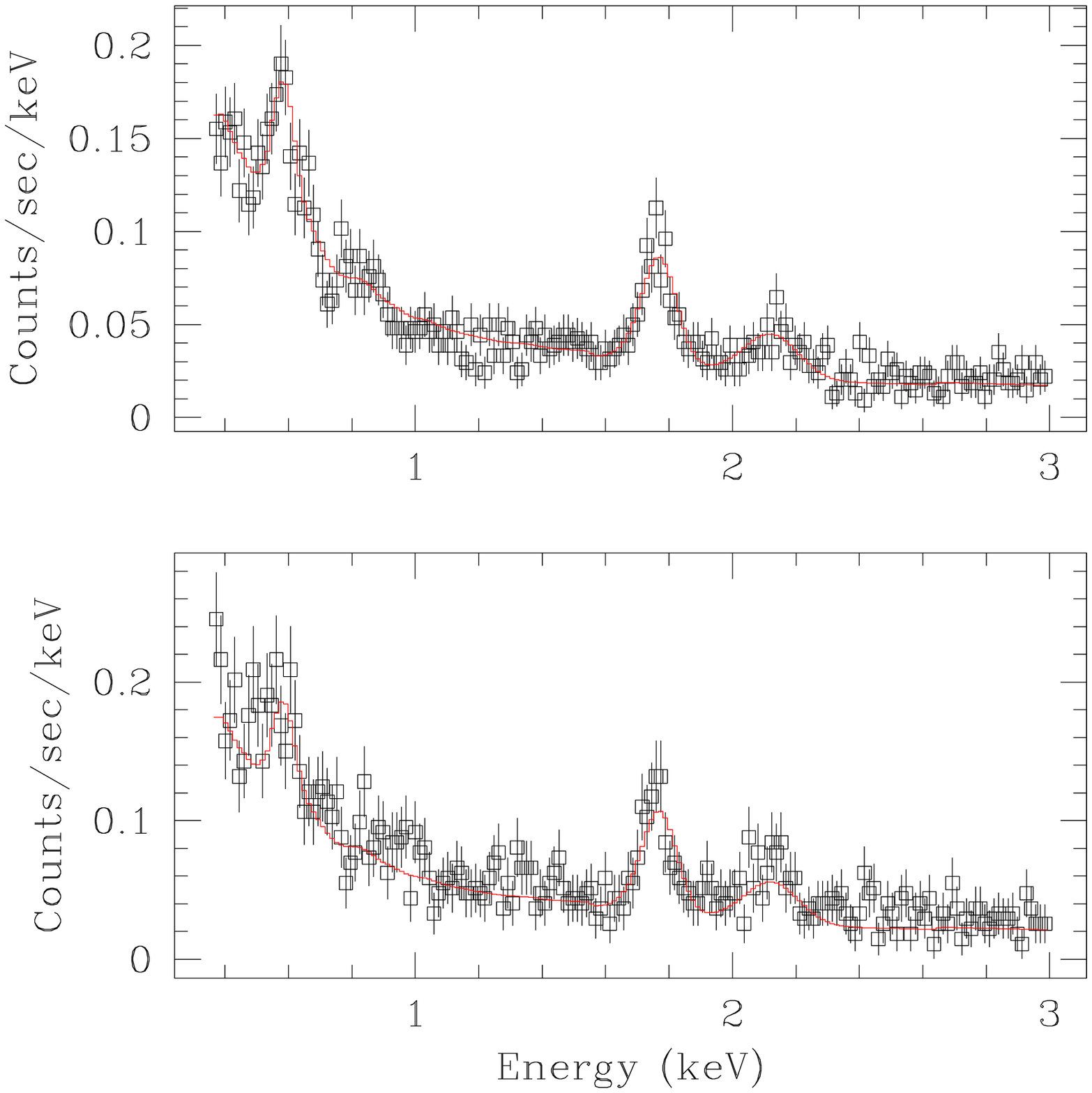}
\figcaption{Spectral fit to on-galaxy data for NGC 3184.  Top: ObsID
804, Bottom: ObsID 1520.}
\end{figure}

\begin{figure}
\epsscale{1.0}
\plotone{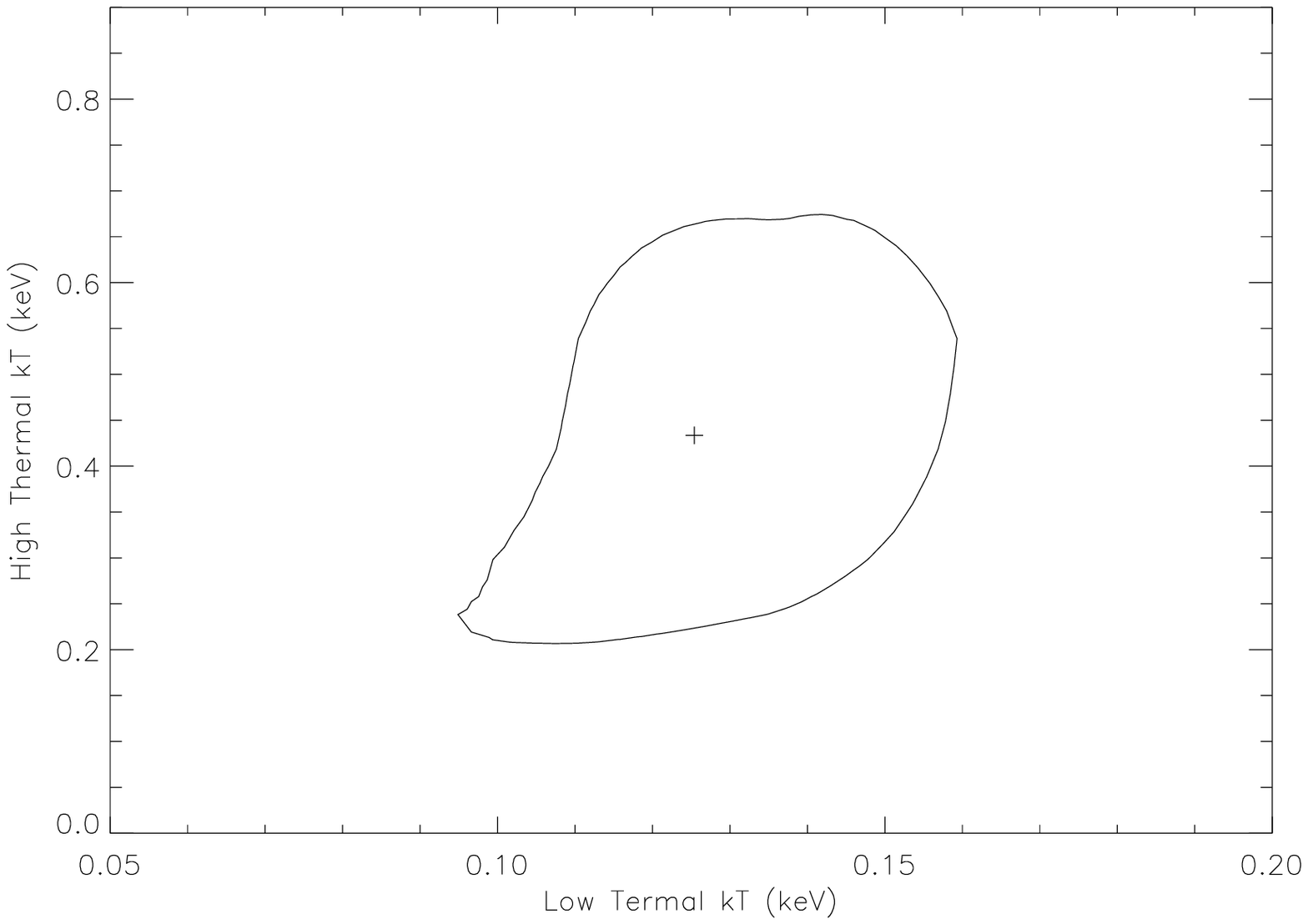}
\caption{The one sigma level in parameter space of low and high
temperatures of the thermal components of the diffuse X-ray emission
from NGC 3184.}
\end{figure}

\end{document}